\documentclass[prb,aps,twocolumn,home,floats,pdflatex]{revtex4}

\usepackage{graphics}
\usepackage{amsmath}
\usepackage{thumbpdf}
\usepackage{citesort}
\usepackage{dcolumn}

\begin{document}

\title{{\rm\small\hfill (submitted to Phys. Rev. B)}\\
On the thermodynamic stability of PdO surfaces}

\author{Jutta Rogal, Karsten Reuter and Matthias Scheffler}

\address{Fritz-Haber-Institut der Max-Planck-Gesellschaft, Faradayweg
4-6, D-14195 Berlin, Germany}

\date{Received: 9 October 2003}

\begin{abstract}
As a first step towards understanding the morphology of PdO crystals
we performed a systematic full-potential density-functional theory 
study of all possible ($1 \times 1$) terminations of the low-index 
surfaces of tetragonal PdO. Applying the concept of
\emph{first-principles atomistic thermodynamics} we analyze the 
composition, structure and stability of these PdO orientations in 
equilibrium with an arbitrary oxygen environment. Within the studied 
subset of ($1 \times 1$) geometries the polar PdO-terminated PdO(100) 
orientation turns out to be surprisingly stable over the whole range 
of experimentally accessible gas phase conditions. Setting up a 
constrained \emph{Wulff construction} within the compiled data set, 
this PdO(100)-PdO facet correspondingly dominates the obtained
polyhedron by far. The real PdO crystallite shape will however
likely be affected by surface reconstructions, which are not
covered by the present study.
\end{abstract}



\maketitle

\section{Introduction}

Metal oxides are compounds of widespread technological interest, and 
one field of application is catalysis, where they can act as the
active material or the (often not that passive) support 
\cite{henrich94,noguera96}. To obtain a microscopic understanding of 
the function of these compounds in such applications it is necessary 
to know their surface atomic structure, which is also 
influenced by temperature and partial pressures in the environment. 
This can be particularly important for oxygen containing environments, 
where the stability of different surface terminations of varying 
stoichiometry may well be anticipated as a function of oxygen in the 
surrounding gas phase. Considering the technological importance of 
oxides the scarcity of such atomic-level information even for 
well-ordered single-crystal surfaces is surprising. The little that is 
known stems almost exclusively from ultra-high vacuum (UHV)
experiments, and is furthermore largely concentrated on some specific 
oxides like the vanadium oxides, rutile (TiO${}_2$) or corundum
(Al${}_2$O${}_3$) \cite{henrich94,noguera96}. For other 
oxides often not even the low-energy surface orientations are firmly 
established, and this also applies for the case of palladium oxide (PdO).

Although PdO is renowned for its high activity in the catalytic 
combustion of methane \cite{burch95,mccarty95,ciuparu01,ciuparu02},
and the involvement of oxidic structures in high-pressure CO oxidation 
reactions at Pd surfaces is now being discussed 
\cite{hendriksen03a,hendriksen03b}, virtually no information about the 
electronic and geometric structure of PdO surfaces is presently 
available, neither from experiment nor from theory. As a first step we 
therefore investigate the surface structure and composition of all 
low-index surfaces of tetragonal PdO in equilibrium with an arbitrary 
oxygen environment using the concept of {\em first-principles
atomistic thermodynamics} \cite{weinert86,scheffler87,kaxiras87,qian88}
based on density-functional theory (DFT) calculations (Section II). 
Lacking any experimental data on surface reconstructions we first
focus on all possible $(1 \times 1)$ terminations and set up a 
constrained Wulff construction for this limited set for the whole
range of experimentally accessible gas phase conditions (Section
IIIB+C). This provides a first data base against which future models of 
reconstructed surfaces may be compared, in particular by how much they 
would have to reduce the surface free energy in order to have
corresponding facets contribute significantly to the real PdO crystal shape.
In addition, relatively high surface free energies and work functions
obtained for $(1 \times 1)$ terminations might point at likely
candidates for surface reconstructions. Interestingly, within the studied 
subset of $(1 \times 1)$ geometries, one termination (PdO(100)-PdO) 
turns out to be much more stable than all others, and correspondingly 
dominates our constrained Wulff polyhedron by far. And this 
although it represents a so-called polar termination, which are traditionally 
dismissed on electrostatic grounds \cite{tasker79,noguera00} (Section IIID).

\section{Theory}

\subsection{Atomistic thermodynamics}

In order to describe the thermodynamic stability of PdO surfaces in an 
oxygen environment we use DFT total-energy calculations as input to 
atomistic thermodynamics considerations 
\cite{weinert86,scheffler87,kaxiras87,qian88,reuter01,reuter03},
which treat the effect of the surrounding gas phase via the contact
with a corresponding reservoir. In equilibrium with such a reservoir
the most stable surface structure in the constant temperature and pressure 
($T$,$p$)-ensemble minimizes the surface free energy, which is defined as
\begin{equation}
\label{equ:gamma}
\gamma(T,\{p_i\})= \frac{1}{A} \left[ \; G - \sum_{i}N_i\mu_i(T,p_i) \; \right] \qquad .
\end{equation}

\noindent
Here $G$ is the Gibbs free energy of the solid with the surface of
area $A$, to which in a supercell calculation with symmetric slabs
both the top and bottom surface contribute equally. $\mu_i(T,p_i)$ 
is the chemical potential of the various species $i$ present in the 
system, i.e. in this case $i =$ Pd, O. $N_i$ gives the number 
of atoms of the $i$th component in the solid. For not too low
temperatures and sufficiently large particles bulk PdO may further 
be assumed as a second thermodynamic reservoir with which the surface 
is equilibrated. This constrains the chemical potentials of O and 
Pd to the Gibbs free energy of bulk PdO, 
$g_{\textrm{PdO}}^{\textrm{bulk}}$ (where the small $g$ denotes the 
Gibbs free energy per formula unit), and allows to eliminate $\mu_{\rm
Pd}$ from eq. (\ref{equ:gamma}). The remaining quantities to be
determined for the calculation of the surface free energy are then the 
chemical potential of the oxygen gas phase, as well as the difference in 
Gibbs free energies of slab and bulk PdO.

The computation of the prior is straight-forward, as $\mu_{\rm O}$ is
completely fixed by the surrounding gas phase reservoir, which may
well be approximated as an ideal gas. Ideal gas laws then relate
the chemical potential to temperature and pressure
\cite{reuter01,janaf}, and we will convert the dependence of the 
surface free energy on $\mu_{\rm O}(T,p)$ in all figures also into the 
more intuitive pressure scales at $T=300$\,K and $T=600$\,K. The
second input to $\gamma(T,p)$, i.e. the Gibbs free energy difference 
of the bulk phase and the slab, receives contributions from changes
in the vibrational and configurational degrees of freedom at the
surface, as well as from the $pV$-term and as leading contribution from
the difference in total energies. From a dimensional analysis, the
$pV$-term may well be neglected \cite{reuter01}. The configurational 
contribution for a system like PdO can further be estimated as below 
5\,meV/{\AA}${}^2$, \cite{reuter03} again negligible for a study that
aims at a first, rather coarse comparison of different $(1 \times 1)$ 
surface terminations. 

The vibrational contribution can be obtained from first-principles 
using the computed phonon density of states (DOS) at
the surface and in the bulk, cf. e.g. ref. \onlinecite{heid00}.
Alternatively, the Einstein approximation to the phonon DOS
can be employed to get an order of magnitude estimate \cite{reuter01}.
Allowing a 50\% variation of the characteristic frequency 
\cite{mcbride91} of each atom type at the surface, the vibrational
contribution to the surface free energy at all considered PdO
terminations stays in this model always within a range of about
10-20\,meV/{\AA}${}^2$ for all temperatures up to $T=600$\,K.
Although this is certainly not a small contribution in general 
anymore, we will nevertheless neglect it in this particular study.
Being interested in a coarse, first screening of the stability of
various PdO terminations, it will become apparent below that even 
a 10-20\,meV/{\AA}$^2$ contribution will not affect the physical
conclusions drawn. If in other studies a higher accuracy is
required, the vibrational term may either be taken into account
by e.g. simplified treatments of the most relevant vibrational
modes \cite{sun03}, or eventually by a full DFT calculation of
the phonon DOS. Here, we will neglect it however, and may with
all other approximations discussed in this section then replace
the difference of bulk and slab Gibbs free energies entering into
the computation of $\gamma(T,p)$ simply by the corresponding
difference of total energies.

\subsection{DFT basis set and convergence}

The DFT total energies that are thus needed are obtained using a
mixture of the full-potential augmented plane wave + local orbitals (APW+lo) and 
the linear augmented plane wave (LAPW) method together with the 
generalized gradient approximation (GGA) for the exchange-correlation 
functional \cite{perdew96} as implemented in the \textsf{WIEN2k}
code \cite{wien2k,sjoestedt00,madsen01}. To simulate the different 
PdO surfaces we use supercells containing symmetric slabs with 7-11 
layers and 12-15\,{\AA} vacuum between subsequent slabs. The 
outermost 2-4 layers are fully relaxed for all surfaces. For all
orientations we have also performed test calculations with thicker
slabs and relaxing more layers, without obtaining any significant
changes to the chosen setup ($\leq$ 3\,meV/\AA$^2$ in the surface 
energy).

The parameters for the mixed APW+lo and LAPW (L/APW+lo) basis set 
are:  $R_{\textrm{MT}}^{\textrm{Pd}} = 1.8$\,bohr, 
$R_{\textrm{MT}}^{\textrm{O}} = 1.3$\,bohr, wave function expansion 
inside the muffin tins up to $l_{\textrm{max}}^{\textrm{wf}}=12$, 
potential expansion up to $l_{\textrm{max}}^{\textrm{pot}}=6$. The
Brillouin zone (BZ) integration is performed using Monkhorst-Pack 
(MP) grids with 6, 8 and 16 \textbf{k}-points in the irreducible
part of the BZ for the (110)/(111), the (100)/(010)/(101)/(011) and
the (001) directions, respectively. The energy cutoff for the plane 
wave representation in the interstitial region is 
$E_{\textrm{max}}^{\textrm{wf}} = 17$\,Ry for the wave function and 
$E_{\textrm{max}}^{\textrm{pot}} = 196$\,Ry for the potential. With 
these basis sets the surface energies of the different PdO surfaces are 
converged within 1-2\,meV/{\AA}$^2$ regarding the \textbf{k}-points and 
3-4\,meV/{\AA}$^2$ regarding $E_{\textrm{max}}^{\textrm{wf}}$. Errors
introduced by the fundamental approximation in the DFT approach, i.e.
our use of a GGA as exchange-correlation functional, will be commented
on below.

\section{Results}

Since there is virtually no atomically-resolved information about 
the structure and composition of crystalline PdO surfaces available 
from the experimental side, we start our theoretical investigation 
from a very basic point of view. That is, to get a first idea about 
the geometric and electronic properties of different PdO surfaces we 
focus here on a rather coarse comparison of the subset of all 
possible ($1\times1$) terminations of the low-index surfaces
of tetragonal PdO.

\subsection{Geometric bulk and surface structure}

\begin{figure}
\scalebox{0.35}{\includegraphics{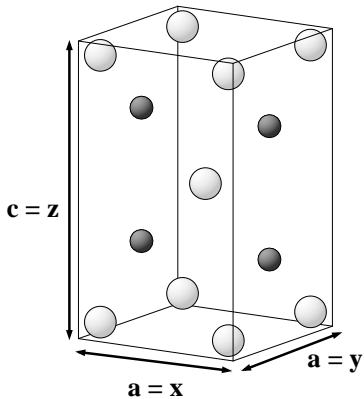}}
\caption{\label{fig:PdOeinheit}
The tetragonal bulk unit cell of PdO.  Small dark spheres indicate
oxygen atoms, large light ones Pd atoms.}
\end{figure}

PdO crystallizes in a tetragonal structure within the space group 
$D_{\textrm{4h}}^9$ \cite{rogers71}. There are two formula units 
of PdO in the tetragonal unit cell with Pd atoms at all corners 
and in the centre, and O atoms at (0, 1/2, 1/4), (0, 1/2, 3/4), 
(1, 1/2, 1/4) and (1, 1/2, 3/4), as shown in Fig. \ref{fig:PdOeinheit}.
All Pd resp. O atoms are equivalent, with each Pd atom planar 
coordinated by 4 oxygen atoms, and each O atom tetrahedrally 
surrounded by 4 Pd atoms. Within our DFT-GGA approach the optimized 
lattice constants of the PdO unit cell are obtained as $a =
3.051$\,{\AA} and $c = 5.495$\,{\AA}, which is in reasonably
good agreement with the experimental lattice constants,
$a_{\rm exp} = 3.043$\,{\AA} and $c_{\rm exp} = 5.336$\,{\AA}
\cite{rogers71}.

Due to the tetragonal structure of the PdO unit cell, there are 
5 inequivalent low-index orientations, each with 2-3 different 
$(1 \times 1)$ surface terminations. The PdO(100) surface 
(parallel to the $yz$-plane) is equivalent to the (010) surface 
(parallel to the $xz$-plane). For this surface direction there 
are two different $(1 \times 1)$ surface terminations, one 
containing only Pd atoms in the topmost layer (PdO(100)-Pd),
and the other Pd as well as O atoms (PdO(100)-PdO). Schematic
drawings of these and all following surface geometries are
shown as insets in Fig. \ref{fig:stability} below. The PdO(001) 
surface is parallel to the $xy$-plane, and there are again two 
$(1 \times 1)$ terminations, one with only Pd atoms (PdO(001)-Pd) 
and one with only O atoms (PdO(001)-O) in the topmost layer.

For the PdO(101) (equivalent to PdO(011)) surface there exist 
three different terminations. One termination is stoichiometric
cutting the stacking of O-Pd${}_2$-O trilayers along the (101) 
direction just between consecutive trilayers (PdO(101)). The other 
two terminate either after the Pd layer (the O-deficient PdO(101)-Pd) 
or have two O layers at the top (the O-rich PdO(101)-O). The 
remaining PdO(110) and (111) directions are on the other hand
characterized by alternating layers of Pd and O atoms along the
surface normal, exhibiting therefore either a completely Pd 
(PdO(110)-Pd, resp. PdO(111)-Pd) or a completely O (PdO(110)-O, 
resp. PdO(111)-O) terminated surface.

Looking at these 11 different $(1 \times 1)$ terminations we 
notice that only one of them is stoichiometric. The other 10
exhibit either an excess of oxygen or palladium atoms, and 
belong thus to the class of so-called \emph{polar} surfaces
\cite{tasker79,noguera00}, the stability issue of which
will be discussed in more detail in Section IIID.

\subsection{Surface free energies}

As mentioned above we want to analyze the stability of these
different PdO surfaces when in contact with an oxygen gas
phase characterized by a given O chemical potential. This 
$\mu_{\textrm{O}}(T,p)$ can experimentally (and assuming that 
thermodynamic equilibrium applies) only be varied within certain 
boundaries. The lower boundary, which will be called the 
\emph{O-poor limit}, is defined by the decomposition of the 
oxide into palladium metal and oxygen. A reasonable upper boundary 
for $\mu_{\textrm{O}}$ on the other hand ({\em O-rich limit}) is 
given by gas phase conditions that are so oxygen-rich, that O 
condensation will start on the sample at low enough temperatures. 
Appropriate and well-defined estimates for these limits are
given by \cite{reuter01}
\begin{equation}
\label{equ:Orange}
\Delta G_f(T = 0 \textrm{K}, p = 0) < \Delta\mu_{\textrm{O}}(T, p_{\textrm{O}_2}) < 0 \qquad ,
\end{equation}

\noindent
where the O chemical potential is referenced with respect to the 
total energy of an oxygen molecule, $\Delta\mu_{\textrm{O}} = \mu_{\textrm{O}} 
- (1/2)E_{\textrm{O}_2}^{\textrm{total}}$, and 
$\Delta G_f(T = 0 \textrm{K}, p = 0)$ is the low temperature
limit for the heat of formation of PdO. Within DFT-GGA we compute 
$-0.88$\,eV for this quantity, which compares well with the 
experimental value of $\Delta G_f^{\rm exp}(T \rightarrow  0 \textrm{K}, 
p = 1 ^{}\textrm{atm}) = -0.97$ eV \cite{CRC95}. To also consider 
the uncertainty in these theoretically well-defined, but approximate 
limits for $\Delta\mu_{\textrm{O}}$, we will always plot the
dependence of the surface free energies also some tenths of eV outside 
these boundaries. From this it will become apparent below that the 
uncertainty in the boundaries does not affect at all our physical 
conclusions drawn.

\begin{figure*}
\scalebox{0.32}{\includegraphics{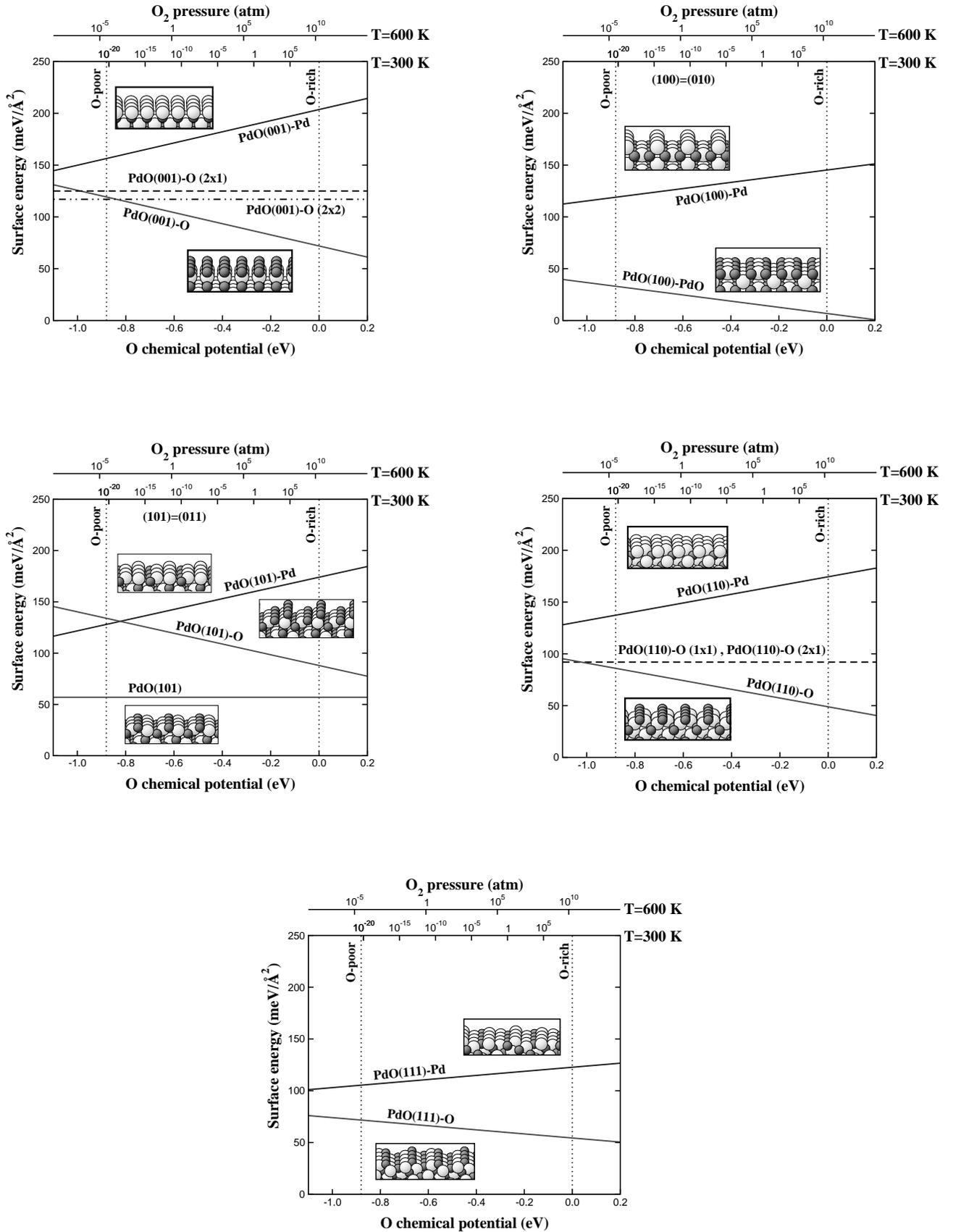}}
\caption{\label{fig:stability}
Surface free energies for the 5 inequivalent low-index PdO surfaces.
Solid lines indicate $(1 \times 1)$ terminations, and dashed lines
larger unit cell reconstructions discussed in Section IIID.
The vertical dotted lines specify the range of $\Delta\mu_{\textrm{O}}$
considered in this study (see text), while in the top two $x$ 
axes the dependence on the O chemical potential has been converted
into pressure scales at $T = 300$ K and $T = 600$ K. The insets
show the surface geometries of the corresponding $(1 \times 1)$ 
terminations (O small dark spheres, Pd large light spheres).}
\end{figure*}

We proceed to show all surface free energy plots
of the eleven discussed $(1 \times 1)$ terminations in Fig. 
\ref{fig:stability}. Terminations with an O excess (deficiency) 
show a negative (positive) slope, i.e. they will become the more
favorable (unfavorable) the more O-rich the surrounding
gas phase is. Comparing the five plots shown in
Fig. \ref{fig:stability} the very high stability of the PdO(100)-PdO
termination is easily recognized. The only other terminations
that exhibit comparably low surface energies are the PdO(101), 
and towards the O-rich limit the PdO(110)-O and PdO(111)-O 
surfaces. All other considered terminations are rather 
high in energy, in particular all those that feature Pd atoms
in their outermost layer. Looking at the energy scale in the
plots we also notice that this stability ordering will not be
affected by the afore discussed 10-20\,meV/{\AA}$^2$ uncertainty
in our surface free energies.

\begin{table}
\caption{\label{tab:gamma}
Surface free energies of all low-index PdO ($1\times1$) terminations
at the oxygen-poor limit, as calculated within the GGA or the 
LDA.  All energies are in meV/\AA$^2$, and the numbers in brackets 
denote the energetic difference with respect to the lowest-energy
PdO(100)-PdO termination.}

\begin{ruledtabular}
\begin{tabular}{l|rr|rr}
Surface 			&
\multicolumn{2}{c}{$\gamma_{\textrm{O-poor}}$}  	&
\multicolumn{2}{c}{$\gamma_{\textrm{O-poor}}$}  \\
termination			&	
\multicolumn{2}{c}{GGA}         &
\multicolumn{2}{c}{LDA}			\\ \hline
\\
PdO(100)-PdO		        &	33    &    (0)
& 59&  (0)	       	\\
PdO(100)-Pd			&	119   &  (+86)				& 170 &(+111)			\\
				&						&				\\
PdO(001)-O			&	119  &	(+86)				& 162 &(+103)			\\
PdO(001)-Pd			&	156  &  (+123)				& 212 &(+153)	 		\\
				&						&				\\
PdO(101)			&	57 &	(+24)				& 86  &(+27)      		\\
PdO(101)-O			&	134 &   (+101)				& 180 &(+121)			\\
PdO(101)-Pd			&	128&	 (+95)
&  173 &(+114)		\\
				&						&				\\
PdO(110)-O			&	86&	 (+53)				& 119 &(+60)			\\
PdO(110)-Pd			&	137&	(+104)				& 173 &(+114)			\\
				&						&				\\
PdO(111)-O            		&         72&     (+39)                		& 109 &(+50)        
     \\
PdO(111)-Pd     	        &        105 &    (+72)
& 143& (+84)            \\
\end{tabular}
\end{ruledtabular}
\end{table}

As already indicated above this uncertainty in the surface energies
does not yet include the error due to the choice of GGA as 
exchange-correlation functional. We have therefore calculated the surface 
free energies of all considered terminations also within the local 
density approximation (LDA) \cite{perdew92}. To set up the supercells 
for these LDA calculations we first optimized the lattice constants 
for PdO bulk, obtaining $a_{\rm LDA} = 2.990$\,{\AA} and $c_{\rm
LDA} = 5.292$\,{\AA}, i.e. values that are as expected slightly
smaller than the afore mentioned experimental lattice constants.
After a full relaxation of the 2-4 outermost layers, the relative
changes in the surface geometries with respect to these bulk values
are found to be very similar to the ones obtained within the GGA.
The resulting absolute surface free energies of the eleven
terminations in the oxygen-poor limit are listed in Table
\ref{tab:gamma} (together with the corresponding GGA values). Although
the absolute values of the LDA surface energies are
30-50\,meV/{\AA}$^2$ higher than within the GGA, the relative
differences between them are within 10-30 meV/{\AA}$^2$. This is exemplified
in Table \ref{tab:gamma} by also indicating the relative energetic
difference with respect to the PdO(100)-PdO termination, which is the
lowest-energy surface in both the LDA and the GGA. As only this
energetic ordering among the considered $(1 \times 1)$ terminations 
enters into the targeted constrained Wulff construction, and we find this
ordering to be similar for calculations with two quite
different exchange-correlation functionals, we expect the DFT
accuracy to be rather high for this system.

\subsection{Constrained Wulff construction}

\begin{figure}
\scalebox{0.40}{\includegraphics{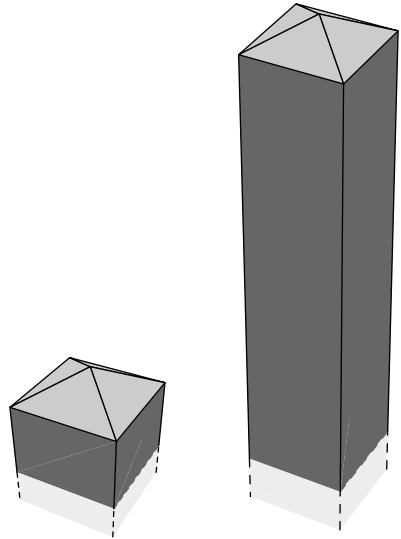}}
\caption{\label{fig:wulff}
{\em Constrained} Wulff construction at the oxygen-poor 
(left) and oxygen-rich limit (right). The construction is 
constrained to the studied $(1 \times 1)$ terminations 
and reflects therefore only the relative energies of 
different PdO orientations, rather than the real PdO crystal
shape, which will most likely be affected by surface
reconstructions. The polyhedra are symmetric with respect 
to the $xy$-plane, and only the upper half is shown
correspondingly.}
\end{figure}

With the results obtained for the surface free energies of the
different $(1 \times 1)$ PdO terminations we set up a constrained 
Wulff construction \cite{wulff01} for a PdO single 
crystal. Since this construction is constrained to reflect
only the studied $(1 \times 1)$ terminations, its intention
is more to compare the relative energies of different 
surface orientations, rather than to really predict the
equilibrium PdO crystallite shape (which will most likely be 
affected by surface reconstructions on which we comment below). 
Since the surface free energies depend on the oxygen chemical
potential, the obtained construction will vary with the conditions 
in the surrounding gas phase. We therefore present in Fig. 
\ref{fig:wulff} the Wulff polyhedra for the two considered 
extremes, i.e. in the O-poor and the O-rich limit. Due to the 
tetragonal symmetry of PdO, the polyhedra are symmetric with 
respect to the $xy$-plane indicated in Fig. \ref{fig:PdOeinheit}, 
and only the upper half of each polyhedron is shown correspondingly. 

As one would already assume from the stability plots in Fig. 
\ref{fig:stability}, the low-energy PdO(100)-PdO surface forms the 
largely dominating facet (rectangular, dark gray facets) at both the 
O-poor and the O-rich limit of $\Delta\mu_{\textrm{O}}$. The other 
triangular, light gray facets are built by the stoichiometric
PdO(101) surface termination. All other investigated terminations 
do not contribute at all to the present construction, since the 
corresponding planes lie outside of the polyhedra and do not cross 
them at any point. The polyhedra in Fig. \ref{fig:wulff} are scaled 
in a way, that the absolute value of the area belonging to the 
PdO(101) termination is the same in the O-poor and O-rich limit,
since the surface free energy of this termination is also constant 
with respect to $\Delta\mu_{\textrm{O}}$. In turn, the remaining surface
area built up by the O-rich PdO(100)-PdO termination increases 
strongly in going from O-poor to O-rich limit. In
the prior limit, this termination forms already 72\% of the whole 
area of the polyhedron, while at the latter limit this fraction has
increased to even 94\%. As would already be expected from the
similar relative energy differences, these Wulff construction
properties are also almost equally obtained within the LDA: again the
PdO(100)-PdO and PdO(101) terminations are the only ones contributing
to the polyhedron, and the PdO(100)-PdO facet covers 66\% (82\%) of
the surface area in the {\rm O-poor} ({\rm O-rich}) limit. Although
such a comparison between two functionals forms no formal proof,
we would therefore assume the obtained shape to be quite
accurate with respect to this DFT approximation.

This does, however, not comprise the major limitation of our study,
given by the restriction to $(1 \times 1)$ terminations.
Surface reconstructions could considerably lower the surface free
energy of any of the PdO surfaces and correspondingly significantly
influence the overall shape of the Wulff polyhedron. Unfortunately, 
to the best of our knowledge no experimental information on such surface
reconstructions is presently available for crystalline PdO. Without any such 
information, not even on the surface periodicity, the phase space
of possible reconstructions is simply too huge to be assertively
screened by todays first-principles techniques alone \cite{lundgren02}. 
Until such information becomes available from experiment, the best we can do
at the moment, is to check by how much such surface
reconstructions would have to lower the surface free energy in order
to give rise to significant changes in the real PdO crystal shape
compared to the constrained construction presented in this study. 
This check is done for the three orientations presently
not contributing to the exterior of the polyhedron, i.e. the (001), 
(110) and (111) facets, and since the surface reconstructions are of
unknown stoichiometry their effect could be different in the
O-poor and in the O-rich limit.

\begin{table}
\caption{\label{tab:reduction}
Minimum energy by which surface reconstructions at the various
facets would have to lower the surface free energy, in order
for the facets to touch the presently obtained constrained
Wulff polyhedron (touching). Additionally, the corresponding 
lowering required for the facet to cover approximately 10\% of 
the total surface area of the polyhedron is listed (10\% area). 
All energies in meV/{\AA}$^2$ (and percent changes) are given 
with respect to the lowest-energy $(1 \times 1)$ termination of 
the corresponding orientation.}

\begin{ruledtabular}
\begin{tabular}{lrr}
Orientation & Touching & 10\% area \\ \hline
\\
Oxygen-poor:\\
(001) & -54 (-45\%) & -67 (-56\%) \\
(110) & -40 (-47\%) & -44 (-51\%) \\
(111) & -7  (-10\%) & -16 (-22\%) \\
\\
Oxygen-rich:\\
(001) &  -7 (-10\%) & -42 (-58\%) \\
(110) & -39 (-80\%) & -41 (-84\%) \\
(111) &  -1 (-2\%)  & -20 (-37\%) \\
\end{tabular}
\end{ruledtabular}
\end{table}

Table \ref{tab:reduction} lists the corresponding values for
each facet, of how much the surface free energy would have to be 
lowered by a reconstruction (with respect to the lowest-energy
$(1 \times 1)$ termination presently considered in our constrained
Wulff construction), such that the facet just touches the current
polyhedron. Additionally given is the corresponding value, required 
to have the reconstruction really contribute significantly
to the total Wulff shape, which we consider to be the case
when the facet covers approximately 10\% of the total surface area. 
From the compiled data in Table \ref{tab:reduction} it seems
that rather massive reconstructions would be necessary (both
in the O-poor and in the O-rich limit) to have 
appreciable (001) and (110) facets in the PdO equilibrium 
crystal shape. A much smaller energy reduction would on the
other hand be required to have a (111) facet contribute. 

\subsection{Stability of PdO surfaces}

\begin{figure}
\scalebox{0.40}{\includegraphics{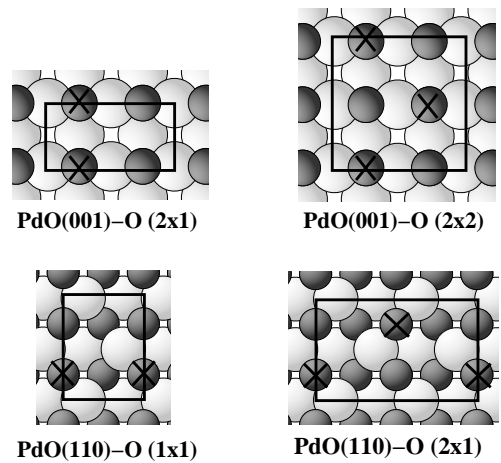}}
\caption{\label{fig:recon}
Unit cells for stoichiometric reconstructions of the PdO(001)-O 
and PdO(110)-O terminations, achieved by simply removing
the oxygen atoms marked with crosses. The two left figures 
show configurations, where the oxygen atoms are taken out 
along a row; the right ones, where the oxygen forms a 
checkerboard pattern.}
\end{figure}

As already mentioned above only one of the 11 possible 
$(1 \times 1)$ terminations, namely the PdO(101), is 
stoichiometric, whereas all others are so-called polar 
surfaces, which are not expected to be stable on electrostatic
grounds \cite{tasker79,noguera00}. For the three different 
surface terminations in the (101) direction this ionic
model certainly complies with our results, since the 
stoichiometric PdO(101) termination turns out much more 
stable than the polar PdO(101)-O and PdO(101)-Pd 
terminations. For the other orientations it is on the other
hand not possible at all to truncate the tetragonal PdO 
structure in a $(1 \times 1)$ cell and achieve charge
neutrality; and one might wonder whether this is the
reason why e.g. the considered (001) and (110) $(1 \times 1)$
terminations exhibit such high surface energies? As most obvious 
guess we therefore performed fully relaxed calculations in
larger unit cells, where we simply removed half the O
atoms in the top layer in order to achieve stoichiometric
terminations. As shown in Fig. \ref{fig:recon} there
are two possibilities for both the (001) and the (110)
orientation to remove these oxygen atoms, either all
along a row or in a checkerboard pattern. The corresponding
surface free energies are drawn as dashed lines in Fig.
\ref{fig:stability}, and are not at all lower than the
corresponding polar $(1 \times 1)$ O-rich terminations.
These results are therefore at variance with the suggestion made
by Ciuparu {\em et al.}, that such a simple removal of O atoms should
lead to charge compensated and thus stable PdO(001) and PdO(110)
surfaces \cite{ciuparu01}.

Apparently, charge neutrality is not the dominant
feature determining the energetic stability of PdO
surfaces, as is also directly reflected in the very 
low surface free energy of the polar PdO(100)-PdO
termination. This points at the most obvious shortcoming
of the electrostatic model, namely the assumption that
all atoms of one species are identical and in the same
charge state, irrespective of whether they are at the
surface or in the bulk. As we had shown before
\cite{reuter01,wang98}, structural and electronic
relaxation at the surface allows for appreciable
deviations from this picture, such that other factors
(like an appropriate excess stoichiometry at O-rich
conditions) might well overrule the polarity issue.

\begin{table}
\caption{\label{tab:workf}
Work functions for the different $(1 \times 1)$ PdO 
surface terminations in eV.}
\begin{ruledtabular}
\begin{tabular}{llll}
O-terminated  		&	$\Phi$ (eV)		&	Pd-terminated	&	$\Phi$ (eV)	\\ \hline
\\
				&				&	PdO(100)-Pd		&	4.0		\\
PdO(001)-O			&	7.9			&	PdO(001)-Pd		&	4.8		\\
PdO(101)-O			&	7.7			&	PdO(101)-Pd		&	4.5		\\
PdO(110)-O			&	7.2			&	PdO(110)-Pd		&	4.4		\\
PdO(111)-O			&	5.9			&
				PdO(111)-Pd		&	4.7 \\ 
\\ \hline
PdO-terminated		&	$\Phi$ (eV)		&	stoichiometric	&	$\Phi$ (eV)	\\ \hline
\\
PdO(100)-PdO		&	6.4			&	PdO(101)		&	5.4		\\
\end{tabular}
\end{ruledtabular}
\end{table}

Still, that there is a different degree of polarity associated
with the different terminations, is nicely reflected in the
corresponding work functions, cf. Table \ref{tab:workf}. While
the work function of the stoichiometric PdO(101) termination
is with 5.4\,eV in a medium range, the work functions of all
Pd-terminated surfaces are about 1.0 - 1.5 eV lower. The ones
of the O-terminated surfaces are on the other hand about
1.0 - 2.5\,eV higher, just as expected from the ionic model.
Even the comparably low work function of the O-terminated
PdO(111)-O surface fits into this picture, as there the layer 
distance to the topmost oxygen atoms is with 0.51\,{\AA} significantly
smaller than for the other orientations, and leads therefore
to a smaller dipole moment. 

\begin{figure}
\scalebox{0.34}{\includegraphics{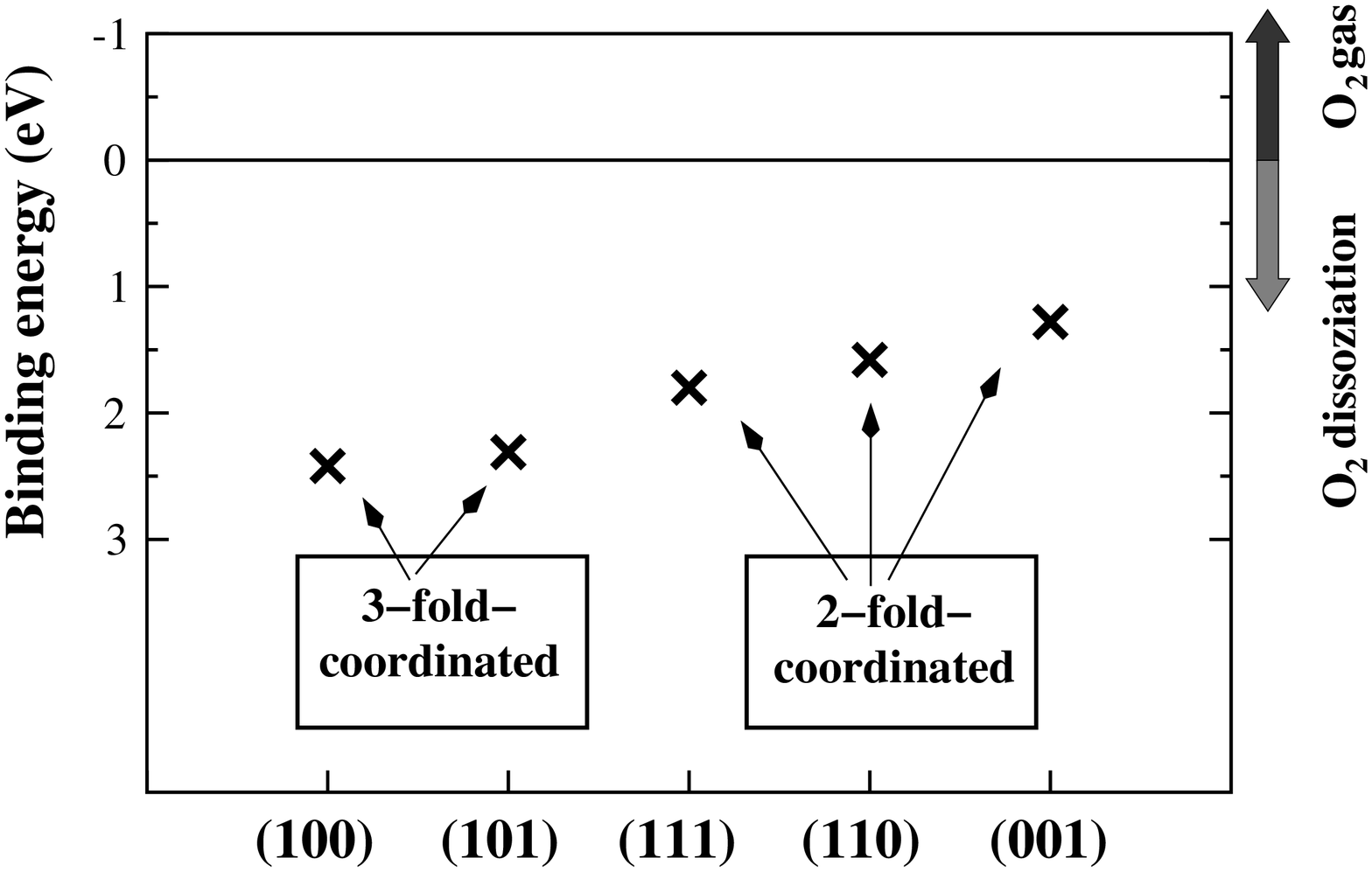}}
\caption{\label{fig:binding}
Binding energies of the topmost oxygen atoms at the most
stable $(1 \times 1)$ termination of the various PdO 
surfaces. The various orientations are sorted along the
$x$ axes with higher surface free energies to the right.
The binding energies are given with respect to a free 
O${}_2$ molecule.}
\end{figure}

Discarding the polarity as a major factor governing the
surface stability, we proceed to correlate the latter with the binding
energy of surface oxygen atoms, as interestingly each most stable 
$(1 \times 1)$ termination of each orientation features oxygen atoms in
the topmost layer, cf. Fig. \ref{fig:stability}. The corresponding 
binding energies with respect to molecular oxygen are shown in 
Fig. \ref{fig:binding}, with positive values indicating that 
O${}_2$ dissociation would be exothermic. The various orientations
are sorted along the $x$ axes with higher surface energies to
the right, revealing a clear correlation. Moreover, there is
also a clear correlation with the coordination of the surface
atoms: The two most stable terminations, namely the ones contributing 
to our constrained Wulff polyhedron, feature three-fold coordinated
surface O atoms and with $\sim 2.5$\,eV rather strong binding
energies. This is followed by the other three orientations,
that do not contribute to the present Wulff shape, exhibiting
only two-fold coordinated oxygen atoms and somewhat lower
binding energies around 1.5\,eV. We therefore conclude that
the stability of the studied subset of $(1 \times 1)$ terminations
seems primarily governed by the openness of the surface 
orientation, i.e. whether its geometric structure offers 
highly-coordinated O binding sites.

\section{Summary}

In conclusion we have calculated the surface free energies of all 
low-index ($1\times1$) PdO terminations in equilibrium with an oxygen 
environment using the concept of \emph{atomistic thermodynamics}.
The PdO(100)-PdO termination exhibits an extraordinarily low surface
energy over the entire range of experimentally accessible gas phase
conditions. Correspondingly, this facet dominates the Wulff polyhedron
constrained to the studied $(1 \times 1)$ terminations by far,
with only the PdO(101) orientation also covering a smaller surface
area. The high stability of these two terminations is largely
connected to their closed geometric structure, allowing a strong 
oxygen binding in highly-coordinated surface sites, while the polarity 
of the non-stoichiometric PdO(100)-PdO termination plays apparently 
only a minor role. The equilibrium shape of a real PdO crystal is 
likely to deviate from the presently obtained constrained Wulff 
polyhedron due to surface reconstructions. Lacking experimental 
information on such reconstructions, only the minimum energy 
lowering required to have corresponding facets contribute to the 
overall crystal shape have been presented.

\section{Acknowledgements}
Stimulating discussions with M. Todorova are gratefully
acknowledged.


\begin{thebibliography}{99}

\bibitem{henrich94}
V.E. Henrich and P.A. Cox, \emph{The Surface Science of Metal Oxides}, 
Cambridge Univ. Press, Cambridge (1994).

\bibitem{noguera96}
C. Noguera, \emph{Physics and Chemistry at Oxide Surfaces}, 
Cambridge Univ. Press, Cambridge (1996).

\bibitem{burch95}
R. Burch and M.J. Hayes, J. Mol. Catal. A \textbf{100}, 13 (1995).

\bibitem{mccarty95}
J.G. McCarty, Catal. Today \textbf{26}, 283 (1995).

\bibitem{ciuparu01}
D. Ciuparu, E. Altman and L. Pfefferle, J. Catal. \textbf{203}, 64 (2001).

\bibitem{ciuparu02}
D. Ciuparu and L. Pfefferle, Catal. Today \textbf{77}, 167 (2002).

\bibitem{hendriksen03a}
B.L.M. Hendriksen, {\em Model Catalysts in Action: High-Pressure Scanning
Tunneling Microscopy}, Ph.D. thesis, Universiteit Leiden (2003);
http://www.physics.leidenuniv.nl/sections/cm/ip/group/
theses.htm\#hendriksen

\bibitem{hendriksen03b}
B.L.M. Hendriksen, M.D. Ackermann, and J.W.M. Frenken, 
({\em private communication}).

\bibitem{weinert86}
C.M. Weinert and M. Scheffler, In: \emph{Defects in Semiconductors}, 
H.J. von Bardeleben (Ed.), Mat. Sci. Forum \textbf{10-12}, 25 (1986).

\bibitem{scheffler87}
M. Scheffler, In: \emph{Physics of Solid Surfaces -- 1987}, J. Koukal 
(Ed.), Elsevier, Amsterdam (1988). M. Scheffler and J. Dabrowski, 
Phil. Mag. A \textbf{58}, 107 (1988).

\bibitem{kaxiras87}
E. Kaxiras, Y. Bar-Yam, J.D. Joannopoulos, and K.C. Pandey, 
Phys. Rev. B \textbf{35}, 9625 (1987).

\bibitem{qian88}
G.-X. Qian, R.M. Martin, and D.J. Chadi, 
Phys. Rev. B \textbf{38}, 7649 (1988).

\bibitem{tasker79}
P.W. Tasker, J. Phys. C {\bf 12}, 4977 (1979).

\bibitem{noguera00}
C. Noguera, J. Phys.: Condens. Matter {\bf 12}, R367 (2000).

\bibitem{reuter01}
K. Reuter and M. Scheffler, Phys. Rev. B \textbf{65}, 035406 (2002).

\bibitem{reuter03}
K. Reuter and M. Scheffler, Phys. Rev. Lett. \textbf{90}, 046103
(2003); Phys. Rev. B \textbf{68}, 045407 (2003).

\bibitem{janaf}
D.R. Stull and H. Prophet, \emph{JANAF Thermochemical Tables}, 2nd 
ed., U.S. National Bureau of Standards, Washington DC (1971).

\bibitem{heid00}
R. Heid, L. Pintschovius, W. Reinhardt, and K.-P. Bohnen,
Phys. Rev. B {\bf 61}, 12059 (2000).

\bibitem{mcbride91}
J. McBride, K. Hass, and W. Weber, Phys. Rev. B \textbf{44}, 5016 (1991).

\bibitem{sun03}
Q. Sun, K. Reuter, and M. Scheffler, Phys. Rev. B \textbf{67}, 205424 (2003).

\bibitem{perdew96}
J.P. Perdew, K. Burke, and M. Ernzerhof, Phys. Rev. Lett. \textbf{77}, 
3865 (1996).

\bibitem{wien2k}
P. Blaha, K. Schwarz, G.K. Madsen, D. Kvasnicka, and J. Luitz, 
\textbf{WIEN2k}, \emph{An Augmented Plane Wave + Local Orbitals
Program for Calculating Crystal Properties}, Karlheinz Schwarz, 
Techn. Universit\"at Wien, Austria (2001). ISBN 3-9501031-1-2.

\bibitem{sjoestedt00}
E. Sj\"ostedt, L. Nordstr\"om, and D.J. Singh, Solid State Commun. 
\textbf{114}, 15 (2000).

\bibitem{madsen01}
G.K.H. Madsen, P. Blaha, K. Schwarz, E. Sj\"ostedt, and 
L. Nordstr\"om, Phys. Rev. B \textbf{64}, 195134 (2001).

\bibitem{rogers71}
D. Rogers, R. Shannon, and J. Gillson, J. Solid State Chem. 
\textbf {3}, 314 (1971).

\bibitem{CRC95}
\emph{CRC Handbook of Chemistry and Physics}, CRC press, Boca Raton FL
(1995).

\bibitem{perdew92}
J.P. Perdew and Y. Wang, Phys. Rev. B \textbf{45}, 13244 (1992).

\bibitem{wulff01}
G. Wulff, Z. Kristallogr. \textbf{34}, 449 (1901).

\bibitem{lundgren02}
E. Lundgren, G. Kresse, C. Klein, M. Borg, J.N. Andersen, M. De Santis,
Y. Gauthier, C. Konvicka, M. Schmid, and P. Varga, Phys. Rev. Lett.
{\bf 88}, 246103 (2002).

\bibitem{wang98}
X.-G. Wang, W. Weiss, Sh.K. Shaikhutdinov, M. Ritter, M. Petersen,
F. Wagner, R. Schl\"ogl, and M. Scheffler, Phys. Rev. Lett. {\bf 81},
1038 (1998).

\end{thebibliography}
\end{document}